\documentclass{article}
\usepackage[whole]{bxcjkjatype}
\usepackage{graphicx}
\usepackage[fleqn]{amsmath}
\usepackage{amssymb}
\usepackage[unicode]{hyperref}

\begin{document}

\title{場の理論の分解学\thanks{雑誌数理科学2012年10月号特集「弦理論と数学」}}
\author{山崎 雅人 (プリンストン大学)}
\date{}
\maketitle

\section{場の理論全体を理解する}

本稿が話題にしたいのは\textbf{場の理論}である．
場の理論というのは量子力学を無限の自由度にしたものである．
その適用範囲は広大であり，素粒子レベルのミクロな構造から
宇宙そのものまで，対象とするスケールも様々である．
物理学手法によって世界を理解する
為の最も基本的なパラダイムの一つであるといっても過言ではない．

それでは，我々は一体どの程度場の理論を理解していると
言えるのだろうか？

普通，場の理論を理解するという時意味されるのは
ある特定の理論を理解することである．例えば，$\phi^4$理論であるとか，
量子電磁力学などはその例である．
しかし，ここで提起したいのはもっとメタな問題
である．
場の理論そのもののなす空間を考えたときに，
その空間のうち我々の既に理解している部分はどれほどだろうか？
もう既にほとんどの場の理論が見つかっているのか，それとも
まだ未知の場の理論の方が多いのか？
また，場の理論の間にどのような関係があり，
場の理論の全体は何か構造を持つのだろうか？

これらの問題に完全な解答を与えることは
極めて難しい問題であるし（最近の
試みとして\cite{Douglas}を参照），
そもそも場の理論を「分類」するというときにそれが正確に何を意味するか
が問題になる．現在場の理論を数学的に完全に厳密に定式化することは
なされていないので，問題自体
厳密に定式化されているわけでもない．
しかし，完全な分類は現在の知識ではできないにしても，
特定のクラスの場の理論に制限することで
何か分かることはないだろうか？

理論のクラスを制限する良い方法は，
次元を下げて自由度を減らしたり，あるいは\textbf{超対称性}のように
高い対称性を課したりすることである．例えば，２次元の固定点を記
述する共形場理論においては，中心荷電と呼ばれている量が1以下の時に
分類が知られている．また，４次元のゲージ理論でも，超対称性が
一番高い場合には
その分類がなされていると考えられている．
しかし，ここで問題にしたいのはより一般の
３，４次元超対称ゲージ理論である．
これらの理論は超対称性によって制限があるものの
制限が分類そのものを与えるほど強くもない．

物理学の勉強をした読者の中には，
問題の答えは既に教科書にあると考える読者もいるだろう．
解析力学の教科書を開くと，そこではいつもラグランジアンや
ハミルトニアンから話は始まるし，場の理論の教科書（の大多数）でも
それに変わりはない．
つまり，
物理学というのは現象を記述する適切な
ラグランジアン・ハミルトニアンを探してくるのが最大の問題であって，
それさえわかればあとは機械的な手続きに従ってそこから出てくる
数学（例えば微分方程式）を解けばいいのである．

では，場の理論のラグランジアンはどんなデータによって指定されるのだろうか？
例として場の理論の教科書によく登場する$\phi^4$理論を考えよう（$\phi$はス
カラー場）．
この場合ラグランジアンは
\begin{equation}
\mathcal{L}=\frac{1}{2}\partial_{\mu} \phi^{\mu} \partial \phi+
\frac{m^2}{2}{\phi^2}+\frac{\lambda}{4!} \phi^4
\end{equation}
である．ただし，$\phi\to -\phi$という対称性を課すことにしたので例えば
$\phi^3$の項はない．
この対称性のもとではラグランジアンに例えば$\lambda_{2n} \phi^{2n} (n\ge 3)$という
項を足すことは許されるので，この理論は無限個のパラメーターを持つようにみえる．
しかし，\textbf{繰り込み群}を考えると状況は変わる．
エネルギースケールを変えると，ラグランジアンのパラメーターは
そのエネルギースケールに応じて変化するが，$\lambda_{2n}$は
エネルギースケールを下げていくと小さくなっていくことが知られている．
一般に
繰り込み群の思想として，低エネルギー領域では理論は
有限個のパラメーターで記述されると考えられている．
場の理論が予言能力を持つのはまさにこのためであり，低エネルギー有効理論は
高いエネルギーの物理の詳細に依存しない．
従って，\textbf{赤外固定点}に議論を限定することにすれば，
ラグランジアンの持つパラメーターは有限であり，
理論にどのような場がいて（この場合はスカラー場$\phi$），さらに
質量は何であるか，またどのように相
互作用するか（この場合は$m, \lambda$）というデータによって理論が決まることになる．
それでは，これが赤外固定点の物理を完全に記述するのだろうか？

しかし，話はいろいろと複雑である．
まず，赤外固定点ではしばしば
結合定数が大きくなり，強く相互作用がおこる．
実際の物性系などで興味を持たれている系の多くが
そうであるし，
ミレニアム問題になっているクオークの閉じこめの
時もそうだ．この場合，ラグランジアンのパラメーターについて
展開する摂動展開は適用できない．
つまり，ラグランジアンがあったからと言って，
そこから実際の物理量に機械的にたどり着けるわけではなく，
例えば格子にのせてスパーコンピューターでシミュレーションすると言った
大変な計算が必要になる｡

ラグランジアンと物理（固定点）との対応は単に複雑なだけではなく，
そもそも１対１ではない．その一つの理由が，場の理論の\textbf{双対性}である．
これは（今の文脈では）高エネルギー領域では二つの別の
ラグランジアンで記述される理論の組が
繰り込み群のもとで同じ赤外固定点に流れていくという現象をさす．
例えば，既に現れた４次元の$\mathcal{N}=4$ゲージ理論では，
電場と磁場を入れ替えるような対称性（S双対性）が存在するので，
結合定数が$g$の理論と$1/g$の
理論は同一視される\footnote{この場合は繰り込みは起こらずラグランジアンレ
ベルで既に固定点上にある．}．
また，4次元の超対称性$\mathcal{N}=1$理論におけるサイバーグ
双対性や3次元のミラー対称性（後に再登場）などは
その良い例である．一般に，二つラグランジアンがあったときに
それが同じ赤外固定点に流れていくのかどうかを判定するアルゴリズムは
知られていない．

問題はそれだけではない．
場の理論の中にはラグランジアンによる記述が存在しない（と信じられて
いる）ものもある．
その良い例が\textbf{６次元の$(2,0)$理論}と呼ばれる共形場理論である．
この$(2,0)$というのは超対称性の数を右向き左向きそれぞれについて
示したものであるが，ここでは超対称性による制約が強いということだけが
重要である．
この理論はもともと弦理論から存在が示唆されたものであるが，
純粋に場の理論における
非自明な固定点としても
理解できるものである．

これらの議論が示すように，ラグランジアンから出発して場の理論を調べる方法
は強力であるが同時に限界もあり，ラグランジアンから実際の物理への写像は
単射でなければ全射でもない．また，ラグランジアンの段階で
双対性を理解するのはかなり非自明であることからも分かるように，
ラグランジアン同士を比較するのも容易ではない．
それならば，場の理論を調べるための何か別の枠組みはあるだろうか？

以下で考えたいのは，（ラグランジアンを持たない）
高次元の場の理論から出発して，それを
それを\textbf{コンパクト化}することによって
低い次元の場の理論を定義するという方法である．
このとき，低次元での場の理論は
コンパクト化する多様体の
選択に依存するので，
一つの理論ではなく理論のクラスを考えることになる．
この理論同士の関係を明らかにするのが目的である．

\section{ゲージ理論を分解する}

具体的にこれから考えたいのは，
先に登場した６次元の共形不変性を持つ$(2,0)$理論である．
例として，
これを２次元多様体$C$および３次元多様体$M$に
コンパクト化することを考えよう．このとき
現れる場の理論をそれぞれ
$T_C$（４次元理論），$T_M$（３次元理論）
と書くことにしよう．ここで，$T$という記号は
理論（theory）からとった．
これらの理論はしばしば紫外領域のデータで記述されるが，
実際に興味のあるのは先に説明したようにどちらかというと
赤外での固定点である．

場の理論のこの定義は
抽象的であるが多様体$C$，$M$に関する数学的な知見が
使えるという利点がある．
もちろん，６次元の理論自体良くわかっていない理論なので，
それをコンパクト化するという操作自体を直接実行することは
出来ない．しかし，６次元理論が存在すること
だけで既に多くのことが従うのである．
以下では\cite{Terashima}に基づいてこれを説明しよう．

多様体の分類で基本的な役割を果たすのは，
多様体を\textbf{分解}する操作である．
ペレルマンによるポアンカレ予想の際に脚光を浴びた
幾何化予想も，三次元多様体の分解に関するものであった．
これをゲージ理論の言葉に読み直すことによって，
ゲージ理論を「分解」したり「貼り合わせ」たりできるはずである．

３次元多様体の分解にはいくつも方法が知られているが，
ここでは３次元多様体$M$を切断することを考えよう．
このとき，切り口にはある２次元多様体$C$が現れるので，
$C$を境界に持つ３次元多様体が現れる．

\begin{figure}[tbp]
\begin{center}
\includegraphics[scale=0.23]{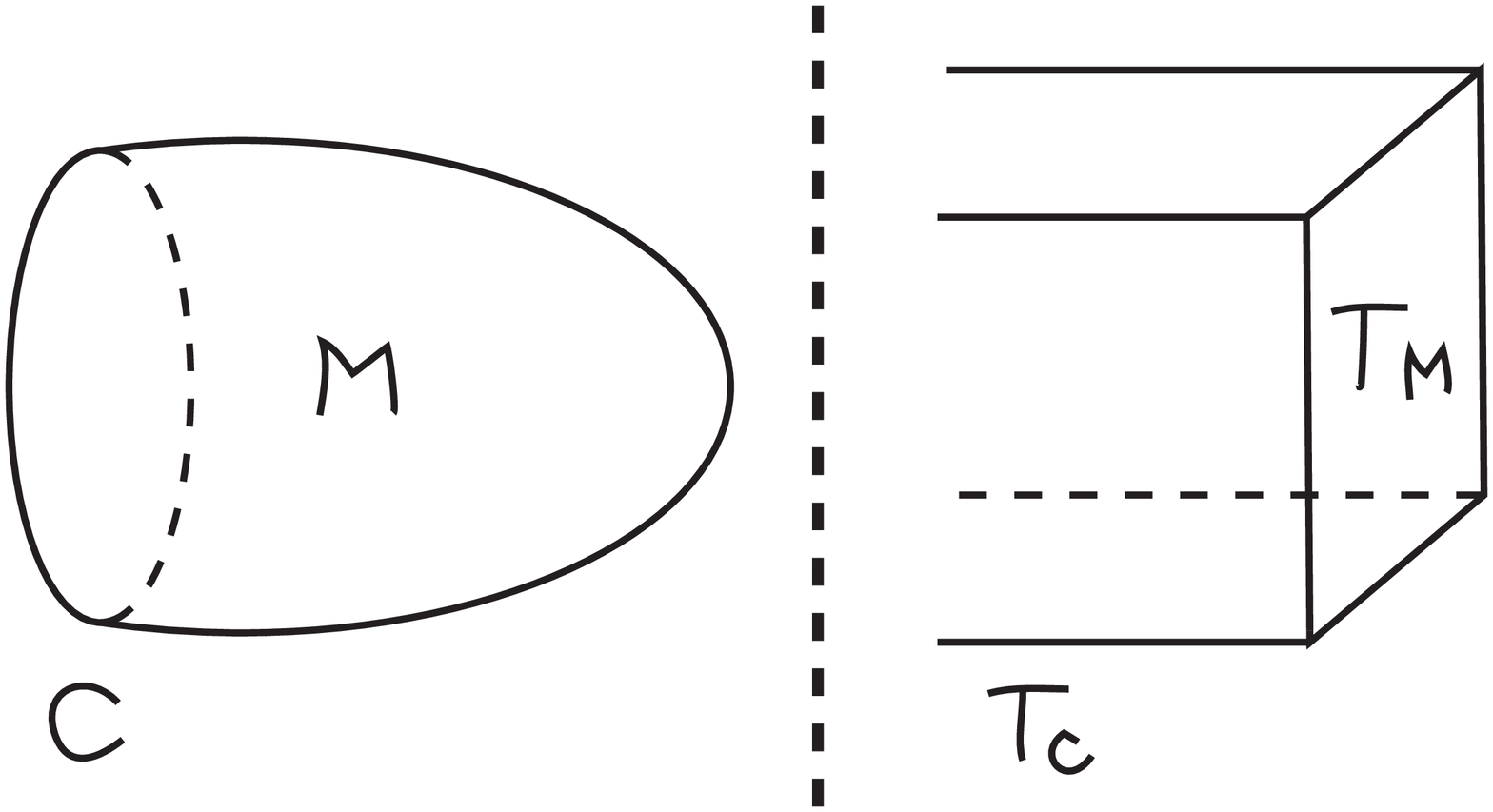}
\end{center}
\caption{境界付きの多様体を考えることは，
ゲージ理論の境界条件を考えることに対応する．}
\label{betweenboundary}
\end{figure}

まず，設定を簡単にするために，
３次元多様体$M$の境界が$C$そのものである状況を
考えることにしよう（図\ref{betweenboundary}）．
このとき，
$M$の分配関数を径路積分で計算したいが，
$M$は
境界$C$を持つので，そこでの境界条件を指定しないと
いけない．理論に存在する全ての場を$\phi$，
境界条件をを$\phi=\phi_0$と書くことにすると，
分配関数は$\phi_0$の（汎）関数になる：
\begin{equation}
Z_M[\phi_0]=\int_{\phi\big|_C=\phi_0} \mathcal{D}{\phi}\,\,
 e^{-S[\phi]}\ .
\label{path}
\end{equation}
これは，$C$に付随した
ヒルベルト空間$\mathcal{H}_C$
の元（波動関数）と見なすことができる：
\begin{equation}
|Z_M\rangle \in \mathcal{H}_C \ .
\label{ZMinC}
\end{equation}
このような定式化は，数学では位相的場の量子論の
公理系\cite{AtiyahSegal}として知られている．
実際，後でみるように我々の設定では$M$上の理論は位相的な理論になっている．

それでは，これをゲージ理論の世界に読み替えよう．
前述したように$T_C$は４次元の理論であり，$T_M$は３次元の理論である．
だから$T_M$の中では$T_C$は四次元$1$を持つように見える．
従って，$T_M$は$T_C$の境界条件とみるのが自然である
（図\ref{betweenboundary}）．

これを次のように書くこともできる．理論$T_C$があると，
それに対応したヒルベルト空間$\mathcal{H}_{T_C}$が
存在する．境界条件を選ぶとは，そのヒルベルト空間
から
一つの元を選ぶことに他ならない：
\begin{equation}
|T_M\rangle \in \mathcal{H}_{T_C} \ .
\end{equation}
これは明らかに\eqref{ZMinC}に対応し，
４次元のゲージ理論が$T_C$が
３次元のゲージ理論$T_C$によって
指定される境界条件を持つことと
解釈できる．

\begin{figure}[tbp]
\centering{\includegraphics[scale=0.23]{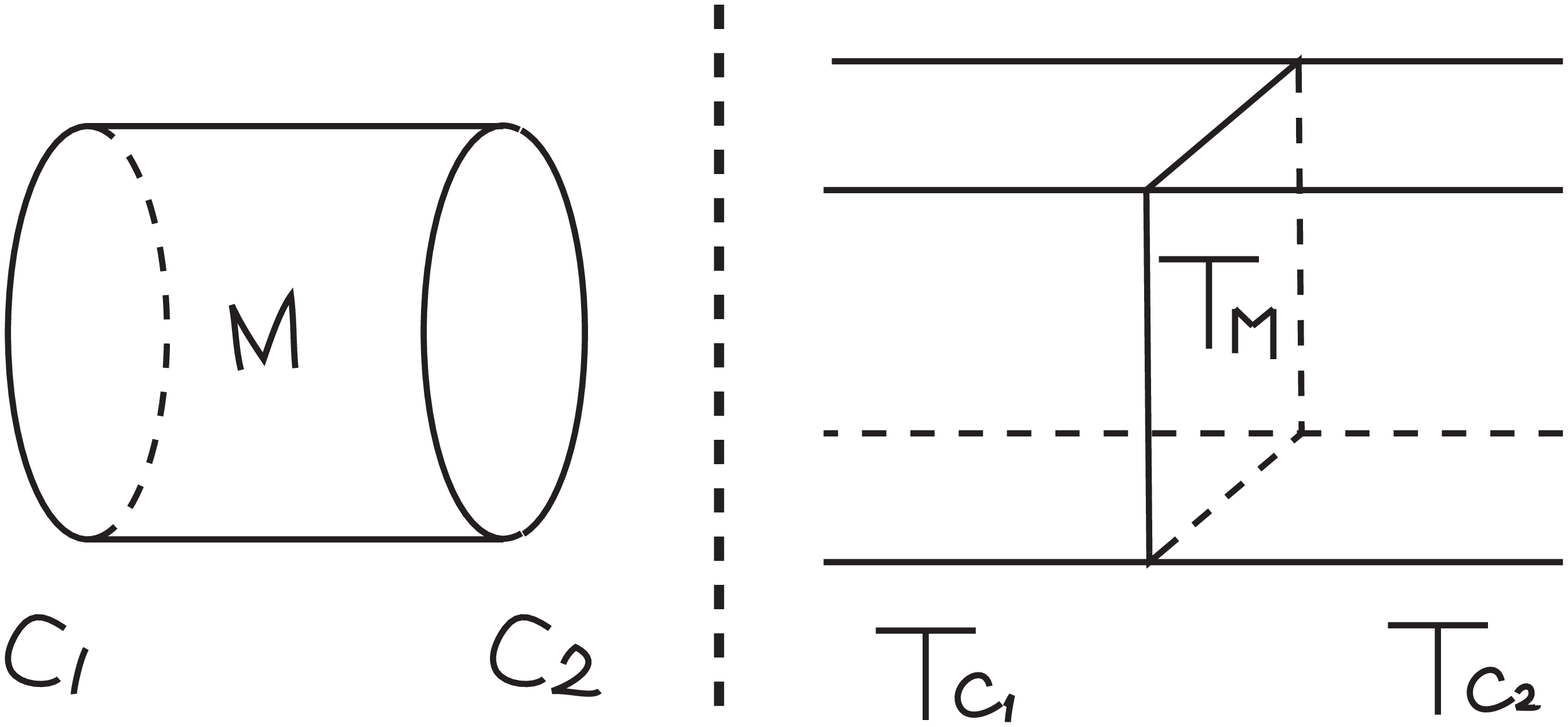}}
\caption{$C_1$と$C_2$をつなぐ多様体$M$は
ゲージ理論$T_{C_1}, T_{C_2}$の間のドメインウォール
対応する．}
\label{withinboundary}
\end{figure}

さて，今度は$M$が二つの境界$C_1$と$C_2$を持つ場合を考えよう（図
\ref{withinboundary}）．
この場合，$M$は$C_1$と$C_2$の\textbf{コボルディズム}であると言われる．
この場合，今度は$Z_M$は二つの境界に付随した
ヒルベルト空間の間の写像と見なすのが
自然である：
\begin{equation}
Z_M \in \textrm{Hom}(\mathcal{H}_{C_1}, \mathcal{H}_{C_2})=\mathcal{H}_{C_1}^*
\otimes \mathcal{H}_{C_2}\ .
\end{equation}
但し，$\mathcal{H}^*$は$\mathcal{H}$の双対
$\mathcal{H}=\textrm{Hom}(\mathcal{H}, \mathbb{C})$であり，
$M$から自然に入る向き付けが$C_1$と$C_2$で異なることから
$\mathcal{H}$かその双対かの違いが生じた．

場の理論での状況は図\ref{withinboundary}
を考えるとわかりやすく，$T_M$は$T_{C_1}$と$T_{C_2}$の間に存在する\textbf{ドメインウォール}に
対応する\footnote{以上の議論は$M$の境界が３つ以上ある場合にも一般化でき，
その場合は$T_M$は複数の$T_C$が集まってできるジャンクション上での
理論と見なすことができる．\label{footref}}．

\begin{figure}[tbp]
\centering{\includegraphics[scale=0.2]{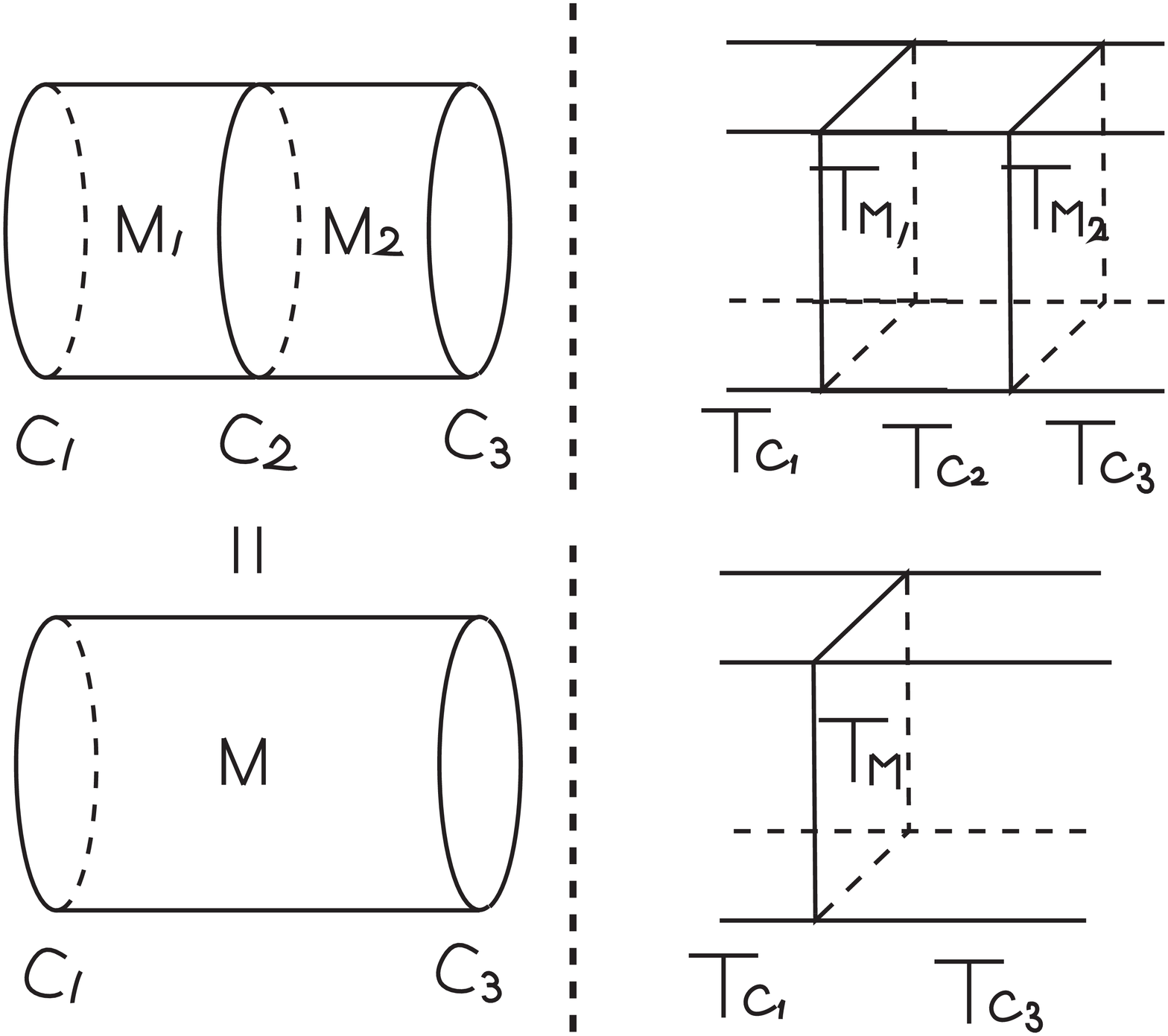}}
\caption{二つのコボルディズムをつなぎ合わせることは
二つのドメインウォールを
重ね合わせることに対応する．}
\label{joinboundary}
\end{figure}

多様体のコボルディズムには
自然な積構造が入る（図\ref{joinboundary}）．
つまり，二つのコボルディズムをつなげてやればよい．
この場合，
\begin{equation}
M=M_1\cup M_2 
\end{equation}
は二つの写像
\begin{equation}
Z_{M_i}\in  \textrm{Hom}(\mathcal{H}_{C_i},
 \mathcal{H}_{C_{i+1}}) \quad (i=1,2)
\end{equation}
の合成を考えることに対応する：
\begin{equation}
Z_{M_1\cup M_2}=Z_{M_1}\cdot Z_{M_2} \in \textrm{Hom}(\mathcal{H}_{C_1},
 \mathcal{H}_{C_3})\ .
\label{compose}
\end{equation}
これは，ゲージ理論の側ではドメインウォールを二つ用意し，それが
重なり合って一つのドメインウォールになる操作だと見なすことができる．
ドメインウォールの間隔は３次元ゲージ理論の
結合定数の逆数を与えるので，ウォールを近づける
操作は赤外固定点に近づくことを意味する．
このようにして多様体のコボルディズム群は
ゲージ理論の境界条件のなす
群に対応する！
与えられた３次元多様体をコボルディズムの積によって分解すると
多様体を幾何的に分解できるが，
今はこれがゲージ理論の分解という物理の問題に翻訳されたことになる．

ここで状況を整理してみよう．
３次元ゲージ理論$T_M$
の立場からすると，それを
４次元ゲージ理論$T_C$の境界条件と考えると言うことは
４次元理論（ゲージ群を$G$とする）との相互作用，ここではゲージ相互作用
の仕方を決めることと同じである．
４次元理論のゲージ場は力学的自由度を持つが，
そのゲージ対称性は３次元理論としては力学的自由度に見えないので
大域対称性に見える．
ドメインウォールは
右と左二つの４次元ゲージ理論と結合しているので，$G\times G$の
大域対称性を持つことになる．
図\ref{joinboundary}にあるようにゲージ理論を貼り合わせるときには
これらの対称性が再度ゲージ化されることになる．

ただし，ここで微妙なのは，しばしば
$G\times G$の
対称性全てがあらわなラグランジアンを書くことは出来ないことである：
例えばある状況では二つの$G$を取り替える操作は３次元のミラー対称性であり，
これは赤外固定点を保つがラグランジアンは大きく変更を受ける．
従って，ゲージ理論を貼り合わせる操作には
一般にラグランジアンにない対称性をゲージ化する必要があり，
一般には$T_M$は
ラグランジアンを持たない理論になる．

こまでの議論をもっと定量的にすることもできる．
具体的には，図\ref{withinboundary}の$M$上の理論の分配関数を計算しよう．
径路積分で境界条件を指定するには，
\eqref{path}のように境界での場の値を
指定してやればよかったが，ここでは$C_1, C_2$に直交する$M$の方向
（水平方向）を時間方向と見な
し，
ハミルトン形式を用いよう．
量子系では場は交換関係を満たしているので，
不確定関係が存在し，全ての場の
値を決めてしまうことは出来ない．そこで，
自由度を位置と運動量に分けて位置についてだけ
条件を課すことになる．$C_1$と$C_2$での位置座標をそれぞれ
$m, \zeta$で現すことにすると
\begin{equation}
Z_M=\langle m| \,\zeta \rangle
\end{equation}
となる．但し，通常の量子力学系ではハミルトニアンによる
時間発展を考える必要があるが，$M$上の理論は位相的なので
ハミルトニアンは自明である．
また，$C$と$C''$での
位置の選択は同じである必要はないので，二つの位置座標
を関係づけるには
正準変換が必要である．それを$\hat{\varphi}_M$で表すと，
\begin{equation}
Z_M=\langle m |\, \zeta\rangle =\langle m |\hat{\varphi}_M|m'\rangle
\end{equation}
と書くことが出来る．
このとき\eqref{compose} 
に相当する主張は
\begin{align*}
&\!\!\!\!\!\!\!Z_M[m, m'']=\int \! [dm'] \, Z_{M_1}[m, m'] Z_{M_2}[m', m''] \\
&=\int \![dm'] \, \langle m | \hat{\varphi}_{M_1} | m' \rangle
 \langle m' | \hat{\varphi}_{M_2} | m'' \rangle \\
&= \langle m | \hat{\varphi}_{M_1}\hat{\varphi}_{M_2} | m'' \rangle
\end{align*}
となる．但し，ここで完全性の式
\begin{equation*}
\int\! [dm]\, |m\rangle \langle m |=1
\end{equation*}
を
用いた．

この分配関数$Z_M$に対応する
超対称ゲージ理論$T_M$側の量が
$S^3$
分配関数である（細道氏の記事を参照）．
この分配関数には紫外発散があるが
正則化することによって
有限な値を持ち，
その答えは先のパラメーター$m, \zeta$に依存する．
ゲージ理論の言葉に翻訳すると，
$m$や$\zeta$というのは
実質量やFIパラメーターと呼ばれる理論のパラメーターであり，
正準変換$\hat{\varphi}_M$
は場の理論を記述する自由度を
取り替える操作であると解釈できる．
例えば，$m$と$\zeta$が互いに位置と運動量を取り替えた関係にある，つまり
$m$と$\zeta$がフーリエ変換の関係にあるとしよう．このとき，
$m$と$\zeta$を取り替える操作は
３次元超対称ゲージ理論のミラー対称性
であり，
それは３次元で粒子の記述と渦糸の記述を取り替える操作である\footnote{この
操作は
物性系，例えば量子ホール効果の物理において重要な役割を果たす．}；
３次元ではゲージ場$A_{\mu}$を考える代わりに
$\partial_{\mu}\phi=\epsilon_{\mu \nu \rho} \partial_{\nu} A_{\rho}$
で関係づいたスカラー場$\phi$を考えてもよく，そのとき
粒子とソリトンとが入れ替わる．
これは４次元で電場と磁場を入れ替える操作の
３次元版である．

より複雑なゲージ理論を得るには，
これまで得たゲージ理論を適宜貼り合わせていけば良く，
一般には分配関数は
\begin{equation}
Z_M=\langle m| \hat{\varphi}_1 \hat{\varphi}_2\ldots \hat{\varphi}_n |m'\rangle 
\end{equation}
の形になる\footnote{この場合$M$はリーマン面上の$S^1$束になっている．
より一般の$M$を得るには注\ref{footref}で述べた３点以上の積を用いる必要が
ある．}．

このようにして出てきた量$Z_M$は
３次元ゲージ理論$T_M$の$S^3$上の分配関数であると同時に
３次元多様体$M$上
の３次元チャーンサイモンズ理論
の分配関数でもあり，$M$の不変量を与える．
特に，$M$が双曲多様体であるときには
$Z_M$は$M$の双曲体積に
量子補正を加えたものであり，
いわば量子\textbf{双曲幾何学}を
記述する\footnote{$M$上のチャーンサイモンズ理論を
境界の２次元面$C$上に制限すると，
理論はリウビユ理論に
なることが知られている．
これの数学的側面については中島氏の
記事を参照．}．

ここまでの議論は抽象的であったが，
$C$に具体的な座標を入れると
演算子$\hat{\varphi}_M$を
具体的に書くこともできる．
このとき，\textbf{量子ダイロガリズム
関数}とよばれる特殊関数が現れ，$Z_M$が
$M$の分解の仕方に依存しないことは
量子ダイロガリズム関数の恒等式によって
保証され，その背後には\textbf{団代数}という
一般的な数学的構造がある（中西氏の記事および\cite{Nagao}を参照）．

興味深いことに，全く同じ構造は
４次元理論$T_C$の中の（超対称性を半分保つ）粒子の束縛状態の研究にも
現れ，それはカラビヤウ多様体の不変量の
\textbf{壁超え}とよばれる
現象の数理と等価である．
つまり，４次元理論の境界条件として現れる
３次元理論の分類が，同じ４次元理論の中の
粒子の束縛状態の分類と対応するのである\cite{Cecotti1}．
さらに，ここで議論した４次元ゲージ理論$T_C$については，
ゲージ群が$SU(2)$の時，
その境界条件のデータが４次ゲージ理論そのものを逆に
決定してしまうと言うことが知られ
ている\cite{Cecotti2}．
このように$T_C$の分類と$T_M$の分類は密接に関係している．

\section{場の理論を超えて}

ここまでは場の理論を
どう捉え直すかという問いに対する
一つの答えをみてきた．
しかし，これまで我々が扱ってきた定量的な量，例えば分配関数は理論一つ一つ
に対して定義された量である．
それでは，場の理論を足しあげて母関数を考えることには
意味はあるだろうか？

このことに関する一つの示唆は重力から得られる．
多様体$M$上の理論はすでに述べたように３次元のチャーン・サイモン
ズ理論であるが，これは
３次元重力と密接な関係にあることが知られている．
重力の理論では時空自体が動的に変化するので
異なる多様体に関する足し上げを考える必要がある．
これは，ゲージ理論の言葉では
ゲージ理論の足し上げに他ならない．

場の理論の標準的な記述に慣れた人間にとって，ゲージ理論を足しあげるという
考え
は突拍子のないもののように思える．
しかし，超弦理論の枠内では次のように考えると
自然に理解できる．ゲージ群として$U(N)$を持つゲージ理論を考えると，
これは超弦理論では$N$枚のブレーンを重ね合わせることで
実現される．
ゲージ理論の立場からすると，$N$というのは固定された数である．
しかし，超弦理論のなかではブレーンは力学的自由度を持つ
対象である．ブレーンを新たに足したり，あるいは
消滅させることは可能であり．
それは$N$の値を増減する．

異なる$N$の値を関係づけるという思想は，
近年可積分系の文脈で具体化された\cite{Nekrasov}．
この設定では2次元の$\mathcal{N}=(2,2)$超対称ゲージ理論$U(N)$で$L$個のフレーバーを持つものを
考える．このとき，この理論の真空は長さ$L$,  上向きスピン$N$個の
スピン$\frac{1}{2}$ XXXスピン鎖で記述されるというのがその主張である．
このスピン鎖においてはスピンを上げたり下げたりする$SU(2)$対称性が存在するので，
これは$N$の値の異なるゲージ理論（の真空）を関係づける$SU(2)$対称性の存在
を意味する\footnote{数学的にはこれは幾何的表現論と呼ばれるものと関係しており，
グラスマニアン多様体のコホモロジーの間に
$SU(2)$の作用を定義できることを意味している．}．
実は，スピン鎖には$SU(2)$のみならヤンギアンとよばれる
より大きな対称性が存在しており，これは場の理論を別の場の理論に
移す対称性が大きなものであり得ることを例示する．

以上，ある種の超対称ゲージ理論を統一的に
理解する一つの視点を，主に位相的場の理論の公理に絡めて
紹介してきた．それは低次元多様体の分解という
豊かな数学と結びつくのみならず，
場の理論のなす空間に作用する新たな対称性を
示唆する．今，場の理論の概念そのものを再度
考え直すことが求められているのではないだろうか？
場の理論とは何であって，
その広大な世界には何が待ちかまえているのか，
その探求は今も続いている．


\end{document}